\documentclass[conference]{IEEEtran}
\IEEEoverridecommandlockouts
\pdfoutput=1
\usepackage{cite}
\usepackage{amsmath,amssymb,amsfonts}
\usepackage{algorithmic}
\usepackage{graphicx}
\usepackage{textcomp}
\usepackage{xcolor}

\usepackage{graphicx}
\usepackage{subfigure}
\usepackage{amsfonts}
\usepackage{amsmath}
\usepackage{xcolor}
\usepackage{booktabs}
\usepackage{booktabs}
\usepackage{multirow}
\usepackage[normalem]{ulem}
\usepackage{url}
\useunder{\uline}{\ul}{}

\def\BibTeX{{\rm B\kern-.05em{\sc i\kern-.025em b}\kern-.08em
    T\kern-.1667em\lower.7ex\hbox{E}\kern-.125emX}}
\begin{document}

\title{HTP: Exploiting Holistic Temporal Patterns for Sequential Recommendation
\thanks{This work was supported by the National Key R\&D Program of China under Grant No. 2020YFB1710200, and the National Natural Science Foundation of China under Grant No.62072136.}
}
\author{{Rui Chen\textsuperscript{1}, Guotao Liang\textsuperscript{1}, Chenrui Ma\textsuperscript{1}, Qilong Han\textsuperscript{*, 1}\thanks{*Qilong Han is Corresponding author}, Li Li\textsuperscript{2}, Xiao Huang\textsuperscript{3}}\\
\textsuperscript{1}\textit{College of Computer Science and Technology, Harbin Engineering University} \\
\textsuperscript{2}\textit{Department of Electrical and Computer Engineering, University of Delaware} \\
\textsuperscript{3}\textit{Department of Computing, The Hong Kong Polytechnic University} \\
\{ruichen, hinleung, cherry1892, hanqilong\}@hrbeu.edu.cn, lilee@udel.edu, xiaohuang@comp.polyu.edu.hk}


\maketitle

\begin{abstract}
Sequential recommender systems have demonstrated a huge success for next-item recommendation by explicitly exploiting the temporal order of users' historical interactions. In practice, user interactions contain more useful temporal information beyond order, as shown by some pioneering studies. In this paper, we systematically investigate various temporal information for sequential recommendation and identify three types of advantageous temporal patterns beyond order, including absolute time information, relative item time intervals and relative recommendation time intervals. We are the first to explore item-oriented absolute time patterns. While existing models consider only one or two of these three patterns, we propose a novel holistic temporal pattern based neural network, named HTP, to fully leverage all these three patterns. In particular, we introduce novel components to address the subtle correlations between relative item time intervals and relative recommendation time intervals, which render a major technical challenge. Extensive experiments on three real-world benchmark datasets show that our HTP model consistently and substantially outperforms many state-of-the-art models. Our code is publically available at \url{https://github.com/623851394/HTP/tree/main/HTP-main}.
\end{abstract}


\section{Introduction}
Recommender systems have been an integral part of many real-world applications, acting as an effective tool for combating information overload, improving user experience, and boosting business revenue. Among various types of recommender systems, sequential recommenders have gained substantial attention for next-item recommendation in different domains, such as e-commerce websites, online video sharing platforms and social media systems. The general idea of sequential recommenders is to explicitly exploit the temporal order of users' historical interactions to predict the next items that users are likely to interact with. There have been a plethora of studies on sequential recommenders based on various sequence models, ranging from Markov chains~\cite{RFT10} to recurrent neural networks (RNNs)~\cite{HKB16,HK18} to self-attention mechanisms~\cite{KM18,LWM20,WLH20} to graph neural networks (GNNs) wherein sequential information is captured during graph construction~\cite{MMZ20,XYY21}.

While sequential recommenders have demonstrated a huge success for next-item recommendation, most of existing research focuses on merely the \textit{temporal order} of users' historical interactions and largely ignores other useful temporal information embedded in historical interactions. Recently, some pioneering works have started to explore other types of temporal information for more accurate recommendations. Time intervals between consecutive interactions are probably the most intuitive temporal patterns to utilize~\cite{ZLW17,JWW20}. Some more recent models such as TiSASRec~\cite{LWM20} and PIMI~\cite{GXY21} further consider time intervals between any pair of interacted items. Another line of research considers the time intervals between the recommendation time and the timestamps of previous interactions~\cite{WZM19,WZM20,WCW20}. Some other works point out that the absolute timestamps, rather than relative time intervals, could also carry beneficial information for recommendation~\cite{WLH20,CHK21}.

\begin{figure}[t]\centering
    \subfigure[Flip flops]{
        \begin{minipage}[t]{0.45\linewidth} 
        \centering
        \includegraphics[width=1\linewidth]{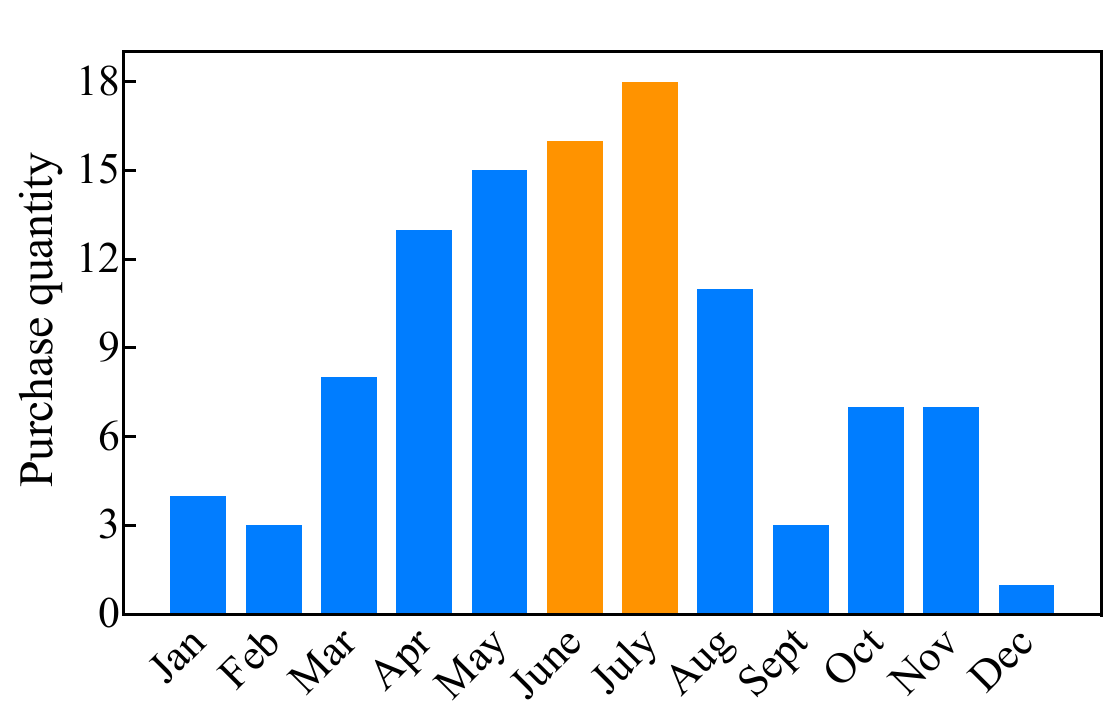}
        \end{minipage}
        }
    \subfigure[Warm jacket]{
        \begin{minipage}[t]{0.45\linewidth} 
        \centering
        \includegraphics[width=1\linewidth]{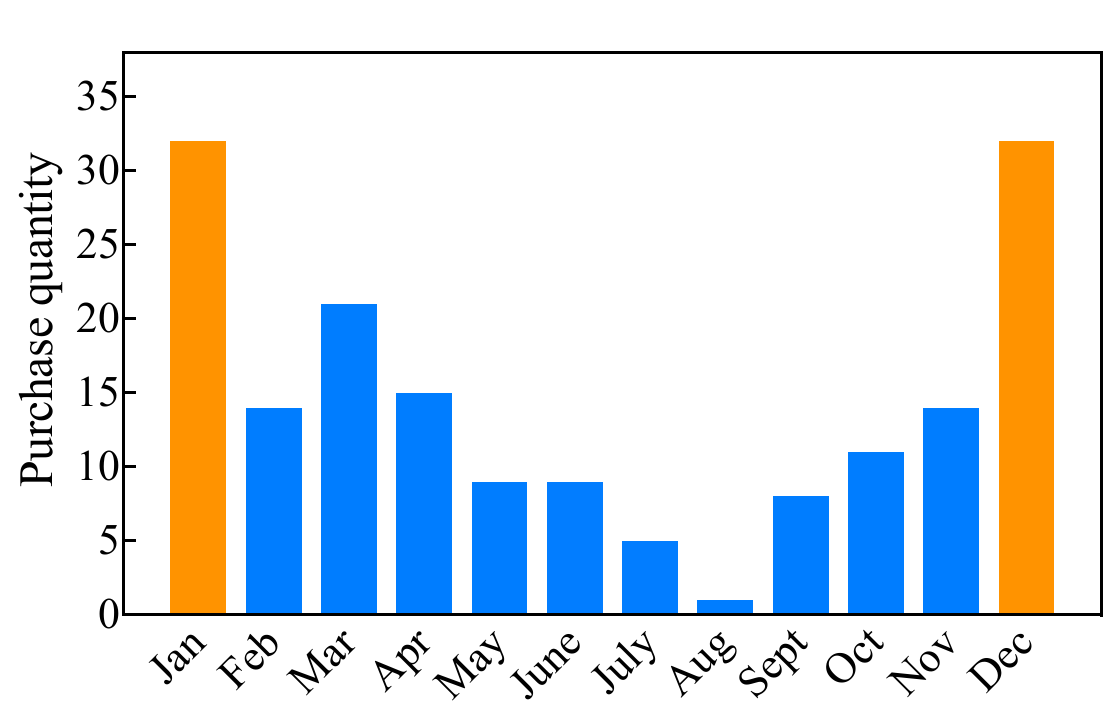}
        \end{minipage}
        }
    \caption{An illustration of item-oriented absolute time patterns learned from Amazon datasets.} 
    \label{fig:motivation}
\end{figure}

Despite the significant improvements achieved by the above efforts, the research on how to make use of diverse temporal information to better next-item recommendation is still in its infancy. In this paper, we provide the first holistic investigation of useful temporal patterns in sequential recommendation. (1) \emph{Absolute time information}: in practice, items are intrinsically associated with absolute time patterns. For example, as illustrated in Fig.~\ref{fig:motivation}, flip flops are more popular in summer while warm jackets are more frequently purchased in cold days. While absolute time information from \textit{the user perspective} was previously studied to learn personal periodic patterns~\cite{WLH20,CHK21}, we are the first to recognize more useful \textit{item-oriented} absolute time patterns. (2) \emph{Relative item time intervals}: considering the relative time intervals between pairs of interacted items (either consecutive items or non-consecutive items) helps unveil their temporal correlations. For example, a user is more likely to purchase a phone case right after purchasing a phone, and users tend to exhibit repeat consumption patterns. 
(3) \emph{Relative recommendation time intervals}: the influence of an interacted item on the next recommendation is related to the relative time interval between its interaction timestamp and the current recommendation time. To the best of our knowledge, \emph{none of the existing models has considered all these temporal patterns}. It is probably due to the technical challenge of capturing the subtle correlations between relative item time intervals and relative recommendation time intervals for next-item recommendation: while it is generally believed that the larger the relative recommendation time interval, the weaker its influence~\cite{WCW20}, the actual influence of a previously interacted item is also compounded by its relative item time interval information. 

In addressing the above challenge, 
we propose a novel \textbf{H}olistic \textbf{T}emporal \textbf{P}attern based sequential recommender, named HTP, to fully leverage all the three temporal patterns. HTP consists of three main components, each of which is designed to handle a specific type of temporal patterns at multiple time granularities (e.g., month, week and day). The absolute time module (ATM) learns each item's global absolute temporal pattern using a data-driven approach. 
The global perspective guarantees the robustness of the learned pattern. The pattern is later used to match the recommendation time so as to indicate a target user' propensity for the item. 
The item time interval module (ITIM) is equipped with multiple time-interval-aware aggregation layers to enhance item representations with temporal periodicity information and establish the temporal correlations among items. To account for the influence of the subtle correlations between item time intervals and recommendation time intervals, the recommendation time interval module (RTIM) not only considers the influence of recommendation time intervals, but also aligns them with the item interval information learned from ITIM. These two aspects are then combined using a nonlinear cascading function, which prevents any single aspect from dominating the recommendation.

We summarize our major technical contributions as follows.
\begin{itemize}
    \item We perform a holistic investigation of useful temporal patterns for sequential recommendation and identify three types of beneficial temporal patterns, including item-oriented absolute time patterns that have not been studied before. 
    \item We propose a novel neural network called HTP to fully leverage the three types of temporal information. In particular, we address the subtle correlations between relative item time intervals and relative recommendation time intervals using a creative cascading design.
    \item Extensive experiments on three real-world benchmark datasets show that our HTP model consistently and substantially outperforms a large number of state-of-the-art models, validating the benefits of leveraging holistic temporal patterns in sequential recommendation.
\end{itemize}

\section{Related Work}
\subsection{Conventional Sequential Recommendation}

Early research on sequential recommendation makes use of Markov chains to mine sequential patterns in historical data. FMPC~\cite{RFT10} models users' long-term preferences by combining first-order Markov chains and matrix factorization for next-basket recommendation. Entering into the era of deep learning, researchers have utilized different neural networks for sequential recommendation. GRU4Rec~\cite{HKB16} is the first model that applies RNNs to session-based recommendation. In sequential recommendation, 
it is important to model both users' long-term preferences and short-term intents. Caser~\cite{TW18} considers convolutional neural networks (CNNs) as the backbone network to embed recent items into an ``image'' so as to learn sequential patterns as local features of the image using convolutional filters. HGN~\cite{MKL19} models long-term and short-term interests via feature gating and instance gating modules. DMAN~\cite{TZL21} segments long behavior sequences into a series of sub-sequences and maintains a set of memory blocks to preserve long-term interests. With the recent emergence of the self-attention mechanism, SASRec~\cite{KM18} captures sequential dynamics based on self attention. There are also other latest results based on self attention~\cite{TZY21}. GNNs are another popular choice for short-term intent modeling. SR-GNN~\cite{WTZ19} models individual session sequences into graph structured data and uses graph neural networks to capture complex item transitions. GCE-GNN~\cite{WWC20} aggregates the global context information and item sequence information in the current sequence through different levels of graph neural networks so as to enhance items' feature representations. 
MGIR~\cite{HZC22} proposes a new direction to enhance session representations by learning from multi-faceted session-independent global item relation graphs. All the above models only consider the temporal order of user interactions and overlook other rich temporal information in sequential recommendation.

\subsection{Time-Aware Sequential Recommendation} 
Time-LSTM~\cite{ZLW17} uses time intervals between consecutive items to model users’ sequential actions based on an LSTM with time gates. SLRC~\cite{WZM19} combines Hawkes processes and collaborative filtering to model temporal dynamics of repeat consumption. Ji \textit{et al.} propose a time-aware GRU network and design a multi-hop memory reading operation to model long-term and short-term preferences~\cite{JWW20}. TiSASRec~\cite{LWM20} utilizes time intervals to model relationships between any two items based on self attention. OAR~\cite{WLH20} captures personal occasions and global occasions based on self attention. CTA~\cite{WCW20} makes use of multiple temporal kernel functions to weigh historical actions' influence. Chorus~\cite{WZM20} takes into consideration both item relations and corresponding temporal dynamics. TASER~\cite{YWC20} leverages absolute time patterns and relative item time intervals to model users' temporal behaviors. Very recently, TimelyRec~\cite{CHK21} learns two specific types of temporal patterns, namely periodic pattern and evolving pattern. 
TGSRec~\cite{FLZ21} jointly models collaborative signals and temporal effects for sequential recommendation.
Despite the recent progress, none of these methods has considered all the temporal patterns we identify.

\section{Problem Formulation}
Let $\mathcal{U}$, $\mathcal{I}$, $\mathcal{T}$ be the user universe, item universe and timestamp universe, respectively. In the setting of sequential recommendation, a user $u \in \mathcal{U}$ is associated with a sequence of interacted items $S_u = [(s_1, t_1), (s_2, t_2), \cdots, (s_{|S_u|}, t_{|S_u|})]$, where $s_i \in \mathcal{I}$, $t_i \in \mathcal{T}$, and $t_i < t_{i+1}$. Each pair $(s_i, t_i)$ indicates that user $u$ interacted with item $s_i$ at timestamp $t_i$. Our goal is to predict the next item that user $u$ might be interested in. In particular, we generate a list of top-$M$ interesting items for $u$ at a future timestamp. 

\begin{figure*}[t]
  \centering
  \includegraphics[width=\linewidth,height=0.4\textwidth]{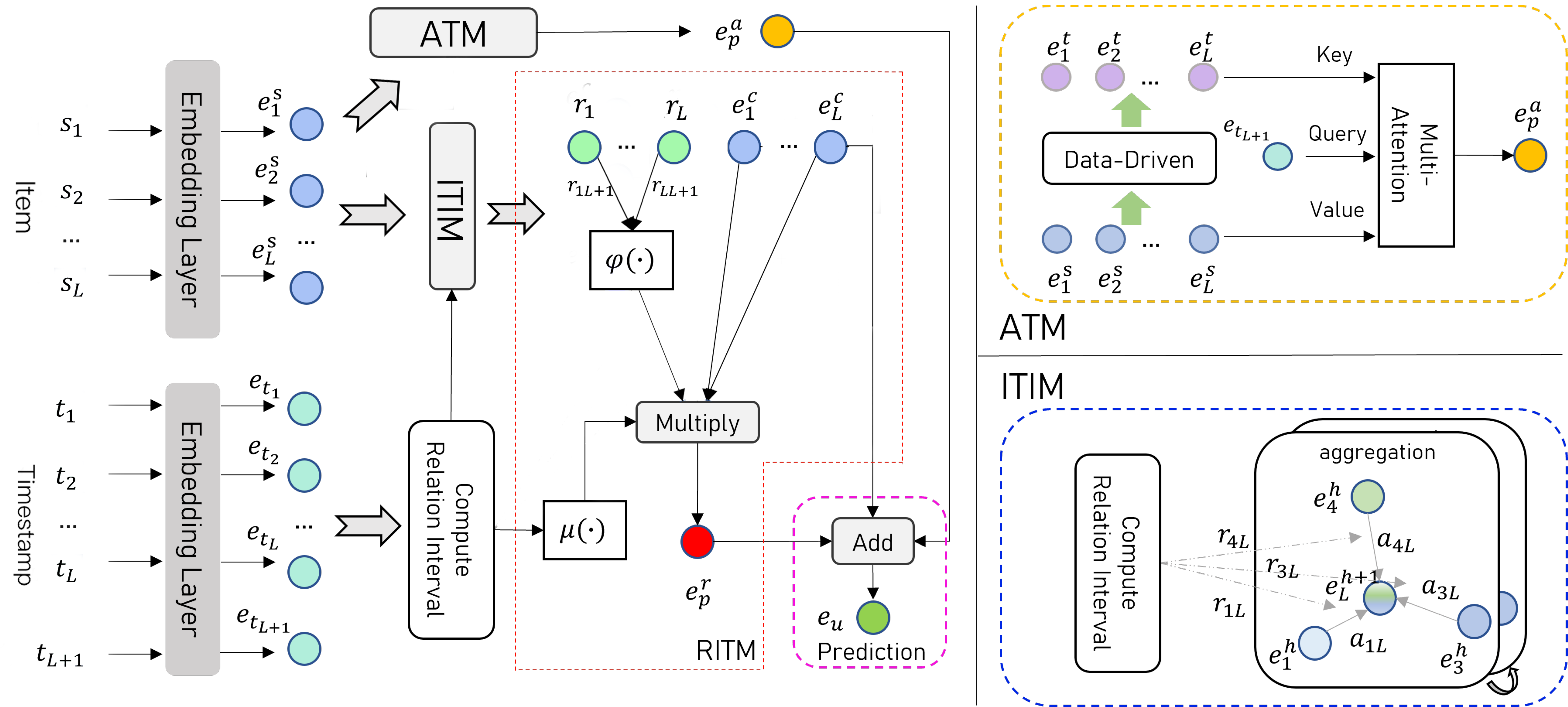}
  \caption{The overall architecture of the proposed HTP model.}
  \label{fig:model}
\end{figure*}

\section{Proposed Method}
We give the overall architecture of the proposed HTP model in Fig.~\ref{fig:model}. It consists of three main components: the absolute time module (ATM), the item time interval module (ITIM) and the recommendation time interval module (RTIM), each of which is designed to tackle a type of temporal patterns. Similar to previous studies~\cite{KM18,LWM20}, if a sequence's length is greater than a predefined value $L$, we truncate it and select the $L$ most recent items; otherwise, we pad the sequence to be of length $L$. 

\subsection{Embedding Layers}
The first step is to map items and timestamps into low-dimensional latent spaces using two different embedding layers. Let $\mathbf{E} \in \mathbb{R}^{{N} \times d}$ denote the item embedding matrix generated by the item embedding layer and $e_i = \mathbf{E}_{i\cdot}$ be the embedding vector of item $i$, where $N$ is the number of items and $d$ is the embedding size. To capture the temporal order among the items in a sequence, we inject a learnable positional embedding~\cite{KM18,VSP17} to each item in the sequence:
\begin{equation}
   e^{s}_{i} = e_i + e^{p}_{i},
\label{eq:position}
\end{equation}
where $e^{p}_{i} \in \mathbb{R}^d$ is the embedding of position $i$ in the sequence. We denote the positional embedding matrix by $\mathbf{M}^p \in\mathbb{R}^{L \times d}$.

To generate the embeddings of timestamps, we consider temporal information at multiple granularities to enrich the temporal representation. In this paper, we consider month-, week- and day-level granularities. The embedding of a timestamp $t_i \in \mathcal{T}$ is computed by
\begin{equation}
    e_{t_i} = e^{m}_{t_i} + e^{w}_{t_i} + e^{d}_{t_i},
\label{eq:time granularity}
\end{equation}
where $e^{m}_{t_i}$, $e^{w}_{t_i}$ and $e^{d}_{t_i}$ are month-, week- and day-level embedding vectors of timestamp $t_i$ extracted from month-, week- and day-level embedding matrices $\mathbf{M}^{m}$, $\mathbf{M}^{w}$ and $\mathbf{M}^{d}$, respectively. Considering timestamp embeddings at multiple levels enables fuzzy matching at various granularities (e.g., two timestamps could be considered similar because they belong to the same month or the same day) and generally improves generalizability.

With the learned item embeddings and timestamp embeddings, we can represent user $u$'s sequence as
\begin{equation}
    S_u = [(e^{s}_{1}, e_{t_1}), (e^{s}_{2}, e_{t_2}), \cdots, (e^{s}_{L}, e_{t_L})],
\label{eq:user sequence embedding}
\end{equation}
which will be used in subsequent modules.

\subsection{Absolute Time Module}
The absolute time module (ATM) learns an item's absolute time pattern using a data-driven approach. We generate the item's absolute time pattern from a global perspective (i.e., consider all users' interactions with this item). The global perspective guarantees the robustness of the learned pattern. Let $T_i$ be the set of interaction timestamps associated with item $i$ from all users. The absolute time pattern of item $i$, $e_{i}^t \in \mathbb{R}^d$, can be formulated as
\begin{equation}
    e_{i}^t = \sum_{t_j \in T_i}{\alpha_{t_j}\,e_{t_j}},
    \label{eq:ATM time aggregation}
\end{equation}
where $\alpha_{t_j} \in \mathbb{R}$ is the normalized weight for a timestamp $t_j \in T_i$. Here $\alpha_{t_j} = \frac{n_{t_j}}{n_i}$, where $n_{t_j}$ is the number of interactions from all users with item $i$ at timestamp $t_j$, and $n_i$ is the total number of interactions from all users with item $i$.

After obtaining items' absolute time patterns, ATM employs a scaled dot-product attention mechanism to measure the extent of matching between the recommendation time (i.e., $t_{L+1}$) and candidate items' absolute time patterns in order to generate a potential ``profile'' of the next item a user is likely to be interested in, which is denoted by $e_p^a \in \mathbb{R}^d$. The general intuition is that the probability of recommending an item should be proportional to the item's popularity at the recommendation time as per its absolute time pattern (e.g., flip flops should be more likely to be recommended in summer but less likely in winter). Formally, $e_p^a$ can be calculated by
\begin{equation}
\begin{split}
    e_{p}^{a} &= \mathrm{Att}(Q, K, V)\\
    &= \mathrm{softmax}\left (\frac{QK^\top}{\sqrt{d}}\right )V,
    \label{eq:ATM attention}
\end{split}
\end{equation}
where $\mathrm{Att}(\cdot)$ is the scaled dot-product attention~\cite{VSP17}, the query $Q$ is the embedding of the recommendation timestamp $e_{t_{L+1}}$, the (key, value) pairs $(K, V)$ are $[(e_{1}^t, e_{1}^s), (e_{2}^t, e_{2}^s), \cdots, (e_{L}^t, e_{L}^s)]$. Recall that $e_i^t$ is item $i$'s absolute time pattern as per Eq.~(\ref{eq:ATM time aggregation}) and that $e_i^s$ is the position-aware item embedding of item $i$ in sequence $s$ as per Eq.~(\ref{eq:position}).

\subsection{Item Time Interval Module}
The goal of the item time interval module (ITIM) is to utilize relative item time intervals to enhance item representations with periodicity information and establish the temporal correlations among different items. To this end, we design a novel time-interval-aware aggregation mechanism that iteratively embeds time interval information via multiple aggregation layers. The $h$th layer updates item embeddings via 
\begin{equation}
\begin{split}
    e_i^{s^{h+1}} &= e_i^{s^h} + \sum_{j = 1}^{K}a_{ij}^h(e_j^{s^h} \cdot \mathbf{W}_1^h),\\
\end{split}
\label{eq:aggregation}
\end{equation}
\noindent where $\mathbf{W}_1^h \in \mathbb{R}^{d \times d}$ is a trainable parameter matrix, and $a^h_{ij}$ gives item $j$'s attention score with respective to item $i$, which we will explain shortly. To remove the noise from less relevant items, we only consider the top-$K$ items that are most relevant to item $i$ in Eq.~(\ref{eq:aggregation}). To properly inject temporal interval information to enhance item embeddings, the attention score function needs a careful design. Inspired by~\cite{CWZ20}, we adopt the time-interval-aware weighted cosine similarity function to calculate $a^h_{ij}$ via
\begin{equation}
    a_{ij}^h = \mathrm{cos}(e_i^{s^h}\cdot \mathbf{W}_2^h, (e_j^{s^h}\cdot \mathbf{W}_3^h+ \mathbf{W}_t^h \cdot r_{ji})),
\label{eq:cos_similarity}
\end{equation}
where $\mathbf{W}_2^h, \mathbf{W}_3^h, \mathbf{W}_t^h \in \mathbb{R}^{d \times d}$ are trainable parameter matrices, and $r_{ji} = e_{t_i} - e_{t_j} \in \mathbb{R}^d$ is the time interval between item $i$ and item $j$ in the sequence. Note that the relative time intervals are calculated using multi-granularity timestamp embeddings, rather than timestamp scalars. The benefits of doing so is observable: (1) the intervals can be naturally compared at different granularity levels; (2) the dense embeddings help the model generalize better.

In addition, we explicitly generate the time interval information for item $i$ as follows:
\begin{equation}
\begin{split}
    r_i^{h+1} &= \sum_{j = 1}^{K}a_{ij}^hr_{ji}.
\end{split}
\label{eq:interval}
\end{equation}
Similarly, only the top-$K$ most relevant items' information is used in Eq.~(\ref{eq:interval}).
After a total of $H$ layers, we use the last layer's output $e_i^{s^{H}}$ as item $i$'s final representation that incorporates time interval information with respect to relevant items, and $r_i^H$ as the time interval information of item $i$. For ease of presentation, we use $e^{c}_{i}$ as a shorthand notation of $e_i^{s^{H}}$ and $r_{i}$ as a shorthand notation of $r_i^H$ in the sequel. 

\subsection{Recommendation Time Interval Module}
The recommendation time interval module (RTIM) is designed to learn the impact of time intervals between the recommendation time and interaction timestamps for next-item recommendation. 
As explained before, recommendation and item time intervals are subtly correlated with each other to influence the next item a user is likely to interact with. Therefore, to evaluate item $i$'s impact on the next recommendation at time $t_{L+1}$, we consider the impact of both the recommendation time interval $r_{iL+1} = e_{t_{L+1}} - e_{t_i} \in \mathbb{R}^d$ and $i$'s time interval information $r_i$ learned from ITIM. 

We first measure the influence of recommendation time interval $r_{iL+1}$ on the next item as follows:
\begin{equation}
\begin{split}
\begin{aligned}
    \varphi(i) &= \frac{\exp(c_i)}{\sum_{j=1}^L \exp(c_j)},\\
    c_i &= r_{iL+1} w,
\end{aligned}
\end{split}
\label{eq:RTIM_time_attention}
\end{equation}
where $w \in \mathbb{R}^{d}$ is a trainable weight vector. Here we introduce $w$ to better make use of the multiple time granularities in $r_{iL+1}$. With $w$, a larger time interval may not always lead to a lower weight. For example, an interaction that happened a long time ago might still have a reasonable impact if the interaction timestamp shares some similarities at different granularity levels (e.g., the same month or same day) with the recommendation timestamp.

Next, we explicitly make use of the ``alignment'' between $r_{iL+1}$ and $r_i$ to depict the next item the user may be interested in:
\begin{equation}
\begin{split}
    \phi(i) &= \frac{\exp (g_i)}{\sum_{j=1}^{L}\exp (g_j)},\\
    g_i &= r_{iL+1}^{\top}\mathbf{W}_rr_i,
\end{split}
\end{equation}
where $\mathbf{W}_r \in \mathbb{R}^{d \times d}$ is a trainable parameter matrix to measure the alignment. As a result, a candidate item that better matches item $i$'s time interval information $r_i$ at time $t_{L+1}$ will have a larger probability to be recommended.


Finally, we introduce a cascading function to combine the weights learned from the above attention mechanisms to delineate the most possible next item by 
\begin{equation}
\begin{split}
\begin{aligned}
    e_{p}^{r} &= \sum_{i = 1}^L \phi(i) \varphi(i) e^c_i.\\
\end{aligned}
\end{split}
\label{eq:RTIM recommended time weight sum}
\end{equation}
It can be seen that our design injects some advantageous nonlinearity to determine the influence of different time interval aspects. In this way, any single aspect is unlikely to dominate the final recommendation. 

\subsection{Fusion and Prediction}
The outputs of the ATM and RTIM modules depict how the next item that a target user will be interested in looks like from distinct perspectives. In addition, as suggested in~\cite{WTZ19}, the latest engaged item $e_{L}^c$ plays an important role in unveiling the user’s most recent interest. We combine them to generate the final profile of the next item:
\begin{equation}
e_{u} = e_{p}^{a} + e_{p}^{r} + e_{L}^{c}.
\label{eq:Fusion}
\end{equation}
While other more advanced fusion strategies might achieve better performance, our experimental results show that Eq.~(\ref{eq:Fusion}) strikes a reasonable balance between model complexity and performance. Finally, we generate the top-$M$ recommendation list by computing the similarity score between a candidate item's embedding $e_o$ and $e_u$ using $r_{o} = e_u e_{o}^\top$ and ranking all candidate items based on their similarity scores.

\subsection{Model Optimization}
We adopt the binary cross entropy loss as the objective function:
\begin{equation}
    - \sum_{(u, t, i, j)\in \mathcal{D}}\log(\sigma(r_{ui}^t)) + \log(1 - \sigma(r_{uj}^t)) + \lambda\Vert \Theta \Vert_F^2,
\label{eq:loss function}
\end{equation}
where $\Theta = \{\mathbf{E}, \mathbf{M}^{w},\mathbf{M}^{d}, \mathbf{M}^{m}, \mathbf{M}^p \}$ is a subset of model parameters, including the item embedding matrix, time embedding matrices and positional embedding matrix, $\Vert \cdot \Vert_F^2$ denotes the Frobenius norm, $\lambda$ is the regularization parameter, and $\sigma(\cdot)$ is the sigmoid activation function. Here $\mathcal{D}$ is the training dataset that combines the positive interaction pair $(u, t, i)$, along with a sampled negative item $j$ that $u$ did not interact with at time $t$. 

\begin{table}[t]
\caption{The statistics of the datasets.}
\begin{center}
\setlength{\abovecaptionskip}{0.3cm}
\setlength{\belowcaptionskip}{0.3cm}
\resizebox{\linewidth}{!}{
\begin{tabular}{@{}cccccc@{}}
\toprule
Datasets       & \hspace{1em} \# of Users  & \hspace{1em} \# of Items  & \hspace{1em} \# of Interactions & \hspace{1em} Avg. Items Per User \\ \midrule
Tafeng        & 16,034 & 8,906   & 644,960       & 40.22 \\
Cloth  & 39,344 & 23,002  & 272,089       & 6.91 \\
Sports & 35,571 & 18,349  & 293,644       & 8.25 \\ \bottomrule 
\end{tabular}
}

\label{tab:datasets}
\end{center}
\end{table}

\begin{table*}[t]
\caption{Performances of different models. The best and second-best results are boldfaced and underlined, respectively. Chorus is not applicable to Tafeng that does not have the co-view information required by Chorus, and thus no results are provided. All improvements are significant with p-value $< 0.05$.}
\resizebox{\textwidth}{!}{
\renewcommand\arraystretch{1.2}
\centering
\begin{tabular}{ccccccccccccccccc} 
\hline
\multicolumn{1}{c|}{\textbf{Datasets}}       & \multicolumn{1}{c|}{\textbf{Metrics}} & \textbf{BPR} & \textbf{FRMC} & \textbf{STAMP} & \textbf{RepeatNet} & \textbf{GRU4Rec} & \textbf{SASRec} & \textbf{GCE-GNN} & \textbf{TiSASRec} & \textbf{OAR} & \textbf{Chorus}  & \textbf{SLRC+} & \textbf{TimelyRec} & \textbf{TGSRec} & \textbf{HTP}    & \textbf{Improv.}          \\ 
\hline

\multicolumn{1}{c|}{\multirow{3}{*}{Tafeng}} & \multicolumn{1}{c|}{NDCG@10}          & 0.3309       & 0.3255        & 0.3394                  & 0.3798             & 0.3112           & 0.3646          & 0.3470           & 0.3878            & 0.3602       & -                  & 0.\textbf{4259}               & 0.3431             & 0.2802          & \uline{0.4165}  & -                      \\
\multicolumn{1}{c|}{}                        & \multicolumn{1}{c|}{HIT@10}           & 0.5149       & 0.5094        & 0.5215                  & 0.5574             & 0.4884           & 0.5623          & 0.5341           & 0.5923            & 0.5585       & -               & \uline{0.6034}                 & 0.5268             & 0.4583          & \textbf{0.6172} & 2.28\%                    \\
\multicolumn{1}{c|}{}                        & \multicolumn{1}{c|}{AUC}              & 0.7987       & 0.7946        & 0.7918                  & 0.8202             & 0.7897           & 0.8292          & 0.8053           & 0.8410            & 0.8259       & -               & \uline{0.8361}                  & 0.8066             & 0.7686          & \textbf{0.8531} & 1.43\%                    \\ 
\hline
\multicolumn{1}{c|}{\multirow{3}{*}{Cloth}}  & \multicolumn{1}{c|}{NDCG@10}          & 0.1174       & 0.1377        & 0.2170                  & 0.1989             & 0.1499           & 0.2034          & 0.2236           & 0.2031            & 0.1776       & 0.2302                     & \textbf{0.2611}  & 0.1583             & 0.1388          & \uline{0.2492}  & -                       \\
\multicolumn{1}{c|}{}                        & \multicolumn{1}{c|}{HIT@10}           & 0.2117       & 0.2351        & 0.3722                  & 0.3316             & 0.2798           & 0.3486          & 0.3625           & 0.3423            & 0.3112       & 0.3818                     & \uline{0.3904}   & 0.2922             & 0.2658          & \textbf{0.4061} & 4.02\%                    \\
\multicolumn{1}{c|}{}                        & \multicolumn{1}{c|}{AUC}              & 0.5856       & 0.6034        & 0.6166                  & 0.6107             & 0.6356           & 0.6860          & 0.6920           & 0.6938            & 0.6664       & \uline{0.6965}                & 0.6709           & 0.6596             & 0.6145          & \textbf{0.7287} & 4.62\%                    \\ 
\hline
\multicolumn{1}{c|}{\multirow{3}{*}{Sports}} & \multicolumn{1}{c|}{NDCG@10}          & 0.2062       & 0.2340        & 0.2530                  & 0.2567             & 0.2416           & 0.3068          & 0.3215           & 0.2845            & 0.2650       & 0.2807                     & \uline{0.3275}   & 0.2360             & 0.1480          & \textbf{0.3346} & 2.16\%                    \\
\multicolumn{1}{c|}{}                        & \multicolumn{1}{c|}{HIT@10}           & 0.3524       & 0.3823        & 0.4203                  & 0.4098             & 0.4224           & 0.4975          & \uline{0.4990}   & 0.4716            & 0.4380       & 0.4496                     & 0.4797           & 0.4121             & 0.2860          & \textbf{0.5335} & 6.91\%                    \\
\multicolumn{1}{c|}{}                        & \multicolumn{1}{c|}{AUC}              & 0.6884       & 0.7048        & 0.6992                  & 0.7153             & 0.7455           & 0.7796          & \uline{0.7846}   & 0.7746            & 0.7559       & 0.7580                     & 0.7434           & 0.7404             & 0.6547          & \textbf{0.8071} & 2.86\%                    \\ 
\hline                     
\end{tabular}
}

\label{tab:experimentalresults}
\end{table*}

\section{Experiments}
In this section, we perform extensive experimental evaluations of the proposed HTP model on three real datasets.

\subsection{Datasets}
In the experiments, we use three real-world benchmark datasets widely used in the literature: \emph{Tafeng}\footnote{https://www.kaggle.com/chiranjivdas09/ta-feng-grocery-dataset}, \emph{Amazon-Cloth}\textsuperscript{\ref{amazon}}, and \emph{Amazon-Sports}\footnote{http://jmcauley.ucsd.edu/data/amazon/links.html \label{amazon}}. The \emph{Tafeng} dataset contains 4-month shopping transactions at the TaFeng supermarket, and the Amazon datasets contain item reviews and metadata in different categories from Amazon. We follow the same procedure in~\cite{KM18,LWM20,WLH20} to preprocess the datasets. In particular, we treat a review or rating as implicit feedback and order the items by timestamps. We keep only the users with at least five ratings or reviews and the items with at least five ratings or reviews. Identical to the settings in~\cite{KM18,LWM20,WLH20}, we use the second last items in the sequences as the validation set for hyperparameter tuning and the last items as the test set. The statistics of the datasets are summarized in Table~\ref{tab:datasets}. 

\subsection{Evaluation Metrics and Baselines}
Following the previous studies~\cite{KM18,LWM20,WLH20}, we adopt two common Top-$M$ metrics, HR@10 and NDCG@10, to evaluate different models. HR@10 counts the percentage of the ground-truth items among the top-10 recommended items. NDCG@10 is a position-aware metric which assigns larger weights to higher positions. To avoid heavy computation on all user-item pairs and accelerate the evaluation process, we follow the strategy in~\cite{KM18,LWM20,WLH20,CHK21}: for each user $u$, we randomly sample 100 negative items and rank them with the ground-truth item. We treat items that user $u$ did not interact with before time $t$ as negative items. To address the sampling bias in evaluation~\cite{KR20}, we also adopt AUC (Area under the ROC Curve) to evaluate different models. We train and evaluate each model five times and report the average results. 

We compare the proposed HTP model with a large number of representative methods given below:
\begin{itemize}
    \item \textbf{BPR}~\cite{RFG09} is a generic recommendation method with Bayesian personalized ranking (BPR). 
    \item \textbf{FPMC}~\cite{RFT10} utilizes Markov chains and matrix factorization for sequential recommendation.
    \item \textbf{STAMP}~\cite{LZM19} models long-term and short-term interests for session-based recommendation, which is a special case of sequential recommendation.
    \item \textbf{GRU4Rec}~\cite{HKB16} employs GRUs to model item sequences for session-based recommendation.
    \item \textbf{SASRec}~\cite{KM18} leverages self attention to model sequentiality without any additional temporal information.
    \item \textbf{RepeatNet}~\cite{RCL19} incorporates a repeat-explore mechanism based on an encoder-decoder structure to improve session-based recommendation.
    \item \textbf{GCE-GNN}~\cite{WWC20} learns item embeddings through different levels of graph neural networks.
    \item \textbf{TiSASRec}~\cite{LWM20} is a state-of-the-art sequential recommender that incorporates item time interval information using self attention. 
    \item \textbf{OAR}~\cite{WLH20} models different shopping occasions by using absolute timestamp information and memorizing the temporal trends of shopping behaviors.
    \item \textbf{Chorus}~\cite{WZM20} takes into consideration both item relations and corresponding temporal dynamics.
    \item \textbf{TimelyRec}~\cite{CHK21} learns two specific types of temporal patterns, namely periodic patterns and evolving patterns.
    \item \textbf{TGSRec}~\cite{FLZ21} jointly models collaborative signals and temporal effects for sequential recommendation.
    \item \textbf{SLRC+}~\cite{WZM19} combines Hawkes processes and collaborative filtering to model temporal dynamics of repeat consumption. Since repeat consumption is removed in the Amazon datasets, we follow the same procedure in~\cite{WZM20} to make it applicable.
\end{itemize}

To make a fair comparison, we do not consider the models whose code is either not publically available or has known bugs that cannot be easily fixed (e.g., TASER~\cite{YWC20}, CTA~\cite{WCW20}, MTAM~\cite{JWW20}). For TGSRec, the results reported in this paper are based on the version after fixing the bugs\footnote{https://github.com/DyGRec/TGSRec/issues/3}.

\subsection{Hyperparameter Settings}

We use a similar method to that in~\cite{LWM20} to determine the maximum length $L$ of input sequences based on the average sequence length, and the maximum length $L$ of \emph{Tafeng}, \emph{Cloth} and \emph{Sports} is set to 50, 15 and 15, respectively. For all datasets, the embedding size $d$ is 50, the number of aggregation layers $H$ in ITIM is 2, the learning rate is 0.0001, and the dropout rate is 0.5. The batch size of \emph{Tafeng}, \emph{Cloth} and \emph{Sports} is set to 256, 1024 and 1024, respectively. The $L_2$ regularization parameter of \emph{Tafeng}, \emph{Cloth} and \emph{Sports} is set to 0.0005, 0.0001 and 0.0001, respectively. The number of the most revelant items $K$ of \emph{Tafeng}, \emph{Cloth} and \emph{Sports} is set to 3, 3 and 2, respectively. For the baselines, we generally follow the settings given in their papers and perform fine-tuning using grid search to identify the optimal performance. We implement our model in PyTorch 1.8 and Python 3.7. All experiments were run on a workstation with an Intel Xeon Platinum 2.40GHz CPU, a NVIDIA Quadro RTX 8000 GPU and 500GB RAM. Our code is publically available at \url{https://github.com/623851394/HTP/tree/main/HTP-main}.

\subsection{Experimental Results}
We present the main experimental results in Table~\ref{tab:experimentalresults}. We can make a few key observations. 
\begin{itemize}
    \item First, our proposed HTP model achieves the best performance in terms of almost all evaluation metrics on all datasets. HTP's improvements are significant with p-value $< 0.05$ and consistent. 
    \item Second, ignoring the temporal order or temporal information cannot achieve reasonable performance for next-item recommendation, as demonstrated by BPR and FRMC. Considering additional temporal information beyond temporal order can bring further performance improvements. For example, SLRC+ and HTP generally perform better than other methods. 
    \item Third, designing a stable model to leverage various temporal information is technically challenging. The existing models that utilize additional temporal information (i.e., TiSASRec, OAR, TimelyRec, Chorus, SLRC+ and TGSRec) cannot always achieve the second-best results. We deem that making use of all three types of temporal patterns is key to capturing complementary temporal information to achieve stable performance.
    \item Finally, from OAR and TimelyRec, we can observe that absolute time information learned from the user perspective is less effective because historical user data are normally highly sparse. In particular, the performance of TimelyRec is less desirable because it needs a large volume of user data (e.g., at least two years of interaction data) to capture sufficient user-level temporal information. This justifies our contribution of considering \textit{item-oriented} absolute time patterns. 
\end{itemize}

\begin{table}[t]
\caption{Benefits of each proposed component. ``w/o'' means without. For example, ``w/o ITIM+RTIM'' means the variant that considers only ATM.}
\setlength{\abovecaptionskip}{0.3cm}
\setlength{\belowcaptionskip}{0.3cm}

\resizebox{\linewidth}{!}{
\begin{tabular}{ccccccc}
\hline
\multirow{2}{*}{Variants} & \multicolumn{2}{c}{Tafeng}        & \multicolumn{2}{c}{Cloth}         & \multicolumn{2}{c}{Sports}        \\ \cline{2-7} 
                          & HR@10           & NDCG@10         & HR@10           & NDCG@10         & HR@10           & NDCG@10         \\ \hline
w/o ITIM+RTIM             & 0.6005          & 0.4003          & 0.3960          & 0.2416          & 0.5158          & 0.3226          \\
w/o ATM                   & 0.5744          & 0.3733          & 0.3598          & 0.2073          & 0.4909          & 0.3024          \\ \hline
w/o month                 & 0.6155          & 0.4100          & 0.4015          & 0.2436          & 0.5248          & 0.3297          \\
w/o week                  & 0.6141          & 0.4146          & 0.4043          & 0.2476          & 0.5218          & 0.3228          \\
w/o day                   & 0.6107          & 0.4079          & 0.4044          & 0.2479          & 0.5285          & 0.3313          \\ \hline
w/o time                  & 0.6025	        & 0.4062           &  0.3935  
& 0.2380	      & 0.5184          & 0.3249          \\ \hline

HTP               & \textbf{0.6172} & \textbf{0.4165} & \textbf{0.4061} & \textbf{0.2492} & \textbf{0.5336} & \textbf{0.3346} \\ \hline
\end{tabular}
}

\label{tab:varians}
\end{table}

\begin{figure*}[t]
  \centering
  \includegraphics[width=\linewidth]{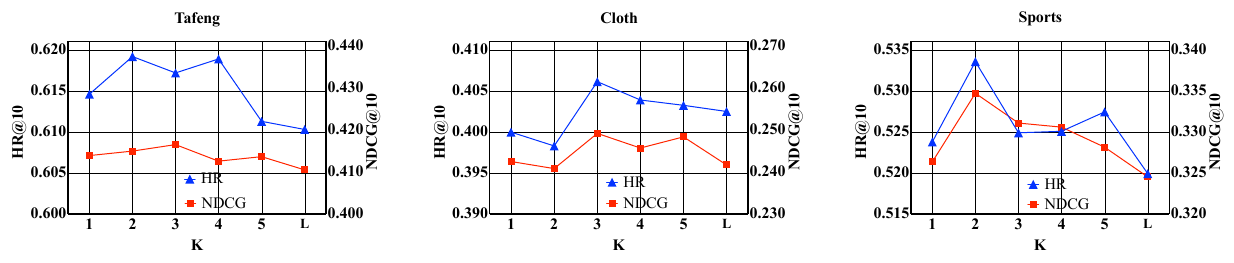}
  \vspace{-5mm}
  \caption{The impact of different numbers of relevant items $K$ on model performance.} 
  \label{fig:K}
\end{figure*}

\begin{figure*}[t]
  \centering
  \includegraphics[width=\linewidth]{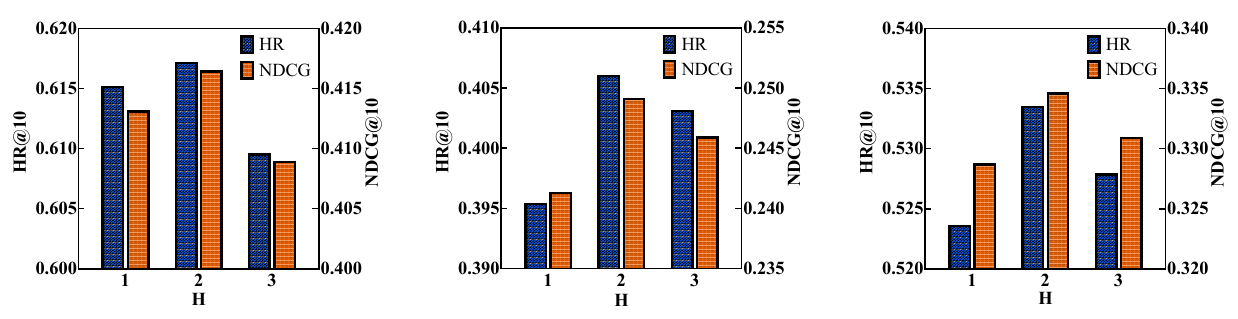}
  \vspace{-5mm}
  \caption{The impact of different numbers of aggregation layers $H$ on model performance. } 
  \label{fig:H}
\end{figure*}

\subsection{Ablation Study}
We perform an ablation study to verify the benefits of having different temporal modules and introducing temporal information in Table~\ref{tab:varians}. We do not separate ITIM and RTIM because by design they jointly address the correlations between item and recommendation time intervals. It can be observed that all proposed modules are advantageous to the final HTP model, which supports the motivation of considering the three types of temporal patterns. ATM is relatively more important. It again justifies the value of learning items' absolute time patterns, which has not been studied before. ITIM and RITM are also indispensable and can further balance periodicity information and time interval information, leading to improved performance. 

Moreover, handling different temporal patterns at multiple time granularities is a valuable idea. We can observe that each time granularity plays a crucial role in capturing items' absolute time patterns. It is interesting to see that different levels of granularities have different extents of importance on different datasets: the month-level granularity is more important for \emph{Cloth}, which confirms our motivation in Fig.~\ref{fig:motivation}~(i.e., flip flops are more popular in summer while warm jackets are more popular in winter). Week-level granularity and day-level granularity are more important for \emph{Sports} and \emph{Tafeng}, respectively. Having all three levels of granularities provides stable performance on different datasets. Last but not least, to further confirm the importance of introducing temporal information, we replace temporal information with position information, and label the results as ``w/o time'' in Table~\ref{tab:varians}. We can observe that substituting temporal information for position information severely spoils the performance. 

\subsection{Hyperparameter Sensitivity} In the last set of experiments, we study HTP's performance with respect to different hyperparameters.

\textbf{Effect of $K$.}  We first study the influence of the number of the most relevant items $K$ in Eq.~\ref{eq:aggregation} and Eq.~\ref{eq:interval}. We vary the value of $K$ in \{1, 2, 3, 4, 5, $L$\} while keeping other hyperparameters unchanged. From Fig.~\ref{fig:K}, we can see that initially increasing $K$ leads to better performance as considering more relevant items allows to bring richer information. However, when $K$ becomes overly large, the aggregation process tends to introduce excessive noise, which degrades the model's performance. In practice, a relatively small $K$ value (i.e., 2 or 3) is able to achieve good performance with small computational cost. 

\textbf{Effect of $H$.} 
The number of time-interval-aware aggregation layers $H$ is a key hyperparameter of HTP. We vary the value of $H$ from 1 to 3 while keeping other hyperparameters unchanged. We present the results in Fig.~\ref{fig:H}. We can see that, with the increase in the number of layers, the model performance improves initially and then declines. With more layers, the ITIM module is able to leverage high-order item correlations to generate more representative item embeddings. However, having too many layers, ITIM will run into an issue similar to the over-smoothing issue in GNNs. In particular, $H=2$ achieves the best performance on all datasets.




\section{Conclusion}
In this paper, we performed a systematic investigation of various temporal patterns for sequential recommendation and identified three types of advantageous temporal patterns, including item-oriented absolute time information, which has not been recognized before. We then proposed a novel holistic temporal pattern based sequential recommendation model, named HTP, which consists of three carefully designed modules, each handling a type of temporal patterns and addressing the shortcomings of previous works. Comprehensive experimental results on three widely used benchmark datasets demonstrate the superiority of our model over a large number of state-of-the-art competitors. 

\bibliography{ref}
\bibliographystyle{unsrt}

\end{document}